\documentclass[aps,prl,twocolumn, titlepage,showpacs]{revtex4}

\usepackage{graphicx}

\usepackage{dcolumn}

\usepackage{bm}

\begin{document}

\title{Accurate particle position measurement from images}

\author{Y. Feng}
\email{yan-feng@uiowa.edu}
\author{J. Goree}
\author{Bin Liu}
\affiliation{Department of Physics and Astronomy, The University
of Iowa, Iowa City, Iowa 52242}

\date{\today}

\begin{abstract}

The moment method is an image analysis technique for sub-pixel
estimation of particle positions. The total error in the
calculated particle position includes effects of pixel locking and
random noise in each pixel. Pixel locking, also known as peak
locking, is an artifact where calculated particle positions are
concentrated at certain locations relative to pixel edges. We
report simulations to gain an understanding of the sources of
error and their dependence on parameters the experimenter can
control. We suggest an algorithm, and we find optimal parameters
an experimenter can use to minimize total error and pixel locking.
Simulating a dusty plasma experiment, we find that a sub-pixel
accuracy of 0.017 pixel or better can be attained. These results
are also useful for improving particle position measurement and
particle tracking velocimetry (PTV) using video microscopy, in
fields including colloids, biology, and fluid mechanics.

\end{abstract}

\narrowtext

\maketitle

\section{INTRODUCTION}
\label{sec:introduction}

Measurement of particle positions from images is important in many
fields, including dusty plasmas \cite{Samsonov:2000, Ivanov:2007},
colloids \cite{Crocker:1996, Kvarnstrom thesis}, fluid mechanics
\cite{Chang:1998}, biology \cite{Selle:2004}, and computer vision
\cite{Shimizu:2005}. Particle positions are generally estimated as
the center of a bright spot of an image. Velocities can also be
calculated from images; two common methods for this are
Particle-Tracking-Velocimetry (PTV) and Particle-Image-Velocimetry
(PIV).

To measure particle positions, an experimenter begins with a
bit-map image. As an example, in Fig.~\ref{experiment images} we
present portions of single video frame from a dusty plasma
experiment. Each bright spot represents an $8~{\rm{\mu m}}$
diameter polymer microsphere illuminated by a $0.633~{\rm{\mu m}}$
helium-neon laser sheet and imaged by a video camera with a Nikon
105 mm micro lens and a bandpass optical filter to eliminate
unwanted light. The lens was focused to generate a sharp image.
The experimental setup is similar to Fig.~2 of
\cite{Samsonov:2000}. Figure~\ref{experiment images}(a) and a
magnified view Fig.~\ref{experiment images}(b) show portions of a
video frame recorded by a cooled 14-bit digital camera (pco1600)
with a $7.4~\rm{\mu m}$ pixel width and a linear response. It was
operated at 30 frames per second with an exposure time of 30 msec.
We should mention that experimental images of particles will
differ, depending on many factors including the type of camera. To
illustrate this point, we present in Fig.~\ref{experiment
images}(c) an enlarged view of a bright spot in a frame recorded
by an analog camera with a nonlinear response corresponding to
gamma = 0.6. (Some cameras are nonlinear with an output intensity
proportional to the input luminance to the power gamma).

In the images in Fig.~\ref{experiment images}, particles fill
several pixels. This spot size may be due, in part, to diffraction
by the particle as well as camera properties such as diffraction
by the camera aperture \cite{aperture book} and imperfect lens
focusing. The spot size cannot be explained merely by geometrical
optics, because the small particle size and magnification would
result in an image smaller than one pixel on the camera detector.

Images have random noise in each pixel. This can arise because of
fluctuations in the camera's sensor and its electronics. Noise in
the experimental image of Fig.~\ref{experiment images}(a) is shown
in Fig.~\ref{histogram} as a histogram of the pixel intensity. The
most prominent feature is the noise peak, corresponding to a large
number of pixels that are relatively dark. This noise peak has an
average value that we term the ``background intensity,'' $I_{bg}$.
The noise peak generally depends only on the camera and the sensor
temperature.

After recording a bit-map image, the experimenter will then use a
computer algorithm to measure the particle position. There are
several methods to do this, including the moment method
\cite{Samsonov:2000, Melandso:2000, Liu:2003, Ivanov:2007,
ImageJ}, which we will study in this paper. Other methods include
fitting a bright spot in the image to a Gaussian \cite{Gai:2001}
or polynomial \cite{Kvarnstrom thesis, Ivanov:2007}, and simpler
methods such as choosing the centroid as the particle center
\cite{Ivanov:2007, ImageJ}. In the moment method, the calculated
particle position is
\begin{equation}\label{1stold}
 {\bf X}_{calc}=\frac{\sum\limits_{k}{\bf X}_{k}I_{k}}{\sum\limits_{k}I_{k}},
\end{equation}
where ${\bf X}_{k}$ is the position and $I_{k}$ is the intensity
of a pixel $k$. The result of Eq.~(\ref{1stold}) is sometimes
called the ``center of mass'' \cite{ImageJ}. When the particle
fills more than one pixel, this calculation yields an estimate of
the particle position with sub-pixel accuracy. Because of the
efficiency and accuracy of the moment method, it is widely used
when analyzing large quantities of data, as might be produced for
example when using a video camera. Fitting methods, which are more
computationally expensive, are often used as well \cite{Gai:2001}.
The centroid method is similar to the moment method except that
the intensity $I_{k}$ of each pixel is replaced with a constant
\cite{Ivanov:2007, ImageJ}.

One application of particle position measurements is the
calculation of particle velocities using PTV. A velocity can be
calculated by subtracting the positions of the same particle in
two different frames and dividing by the time interval between
frames. This method differs from PIV \cite{Westerweel:1997}, where
velocities are calculated at regular gridpoints, not for specific
particles.

Errors in the calculated particle position arise from multiple
sources, including random noise in each pixel and also from the
finite spatial resolution of the pixels on a camera sensor. When
an image is recorded by sampling it with a finite number of
pixels, some information about the intensity profile is lost, and
this can cause a type of systematic error known as pixel locking
or peak locking. The total error in the calculated position will
be due to a combination of these effects, not just random noise or
pixel locking by itself.

In this paper, we seek to minimize the total error, and doing this
will require that we understand the contribution of pixel locking.
Our goal is to aid the experimenter in making optimal choices, in
both hardware and software, to minimize the total error.

\section{PIXEL LOCKING}
\label{sec:pixellock}

Pixel locking, also known as peak locking, is an artifact where
calculated particle positions tend to be concentrated at certain
favored locations relative to pixel edges, such as the center or
edges of a pixel. It is different from random errors, which do not
result in favored positions for particles. To understand pixel
locking, consider a particle whose image fills only a single
pixel. In this case, the sum in Eq.~(\ref{1stold}) would have only
a single term, and the position would be assigned to the exact
center of that pixel. If the particle's image instead fills two
pixels with equal intensity, the position will be assigned to the
midpoint of a pixel edge. The pixel center and midpoints of pixel
edges are examples of favored positions that are found to occur
even when the particle's image fills several pixels
\cite{Nosenko:2006}.

The scientific literature for pixel locking includes many papers
where PIV is used to measure velocities. In the early 1990s, the
PIV method was tested to demonstrate their sub-pixel accuracy for
particles flowing along with a fluid \cite{Willert:1991,
Prasad:1992}. For specific applications of PIV, pixel locking has
been studied by other authors as well \cite{Gui:2002,
Christensen:2004, Angele:2005, Nabach:2005}. In comparison to PIV,
the literature for PTV includes fewer studies of pixel locking,
e.g.  \cite{Nosenko:2006, Melzer:2006}. Because of this, some
users of PTV, including until recently the authors of this paper,
were unaware of pixel locking and the problems it can cause. In
addition to PTV, computer vision is another important area where
pixel locking is recognized as a problem in measuring positions
\cite{Shimizu:2005, Stein:2006, Nehab:2006}.

To detect pixel locking, we use sub-pixel maps as a diagnostic
tool. A sub-pixel map shows all the calculated particle positions
relative to pixel edges, and it is drawn in a small box having the
size of one pixel. To prepare a sub-pixel map, we begin with a
graph of calculated positions of $N$ particles, as illustrated in
Fig.~\ref{sub-pixel map}(a), then plot the fraction parts of these
positions in the small box, yielding the sub-pixel map in
Fig.~\ref{sub-pixel map}(b). In Fig.~\ref{sub-pixel map}(c) we
present an actual sub-pixel map calculated from a bit-map image by
an analog camera in a dusty plasma experiment. The signature of
pixel locking can be identified in general by concentrations of
calculated particle positions at favored positions. These favored
positions can vary, depending on both hardware and software, but
they commonly include the center or edges of a pixel, as in
Fig.~\ref{sub-pixel map}(c). Sub-pixel maps are therefore very
useful for detecting pixel locking. Other authors have used
similar graphs, where the calculated positions have been binned
and plotted as a histogram \cite{Shimizu:2005, Stein:2006,
Nehab:2006}.

\section{MOMENT METHOD}
\label{sec:momentmethod}

The algorithm we optimize in this paper, the moment method, has
two main steps. The first step is the selection of pixels that
belong to each particle in the image. The second step is the
calculation of position as an intensity-weighted moment of pixel
positions.

In the first step, the selection of pixels, the user begins by
choosing a threshold $I_{th}$. The gray-scale image is replaced by
a black-and-white image, where pixels brighter than $I_{th}$
become black, and all others become white. The choice of the
threshold is important for several reasons \cite{Melandso:2000},
as we will discuss later. Next, the boundaries for individual
particle images are determined. There are several algorithms for
selecting boundaries. We have examined several codes that use the
moment method, and we found that the only difference is the
algorithm for selecting boundaries. We will consider three
algorithms, which we distinguish by the corresponding codes we
will test. All three of these codes are well tested, and they
generate reliable results from experimental images. In one
algorithm, the boundary is selected to be a polygon that encloses
only contiguous pixels brighter than the threshold,
Fig.~\ref{boundaries}(a). This algorithm is used in the freely
available ImageJ \cite{ImageJ} code. The other two algorithms
select a boundary that is a rectangle. In Code A, the boundary is
the smallest rectangle that encloses all the contiguous pixels
above the threshold \cite{Code A}, Fig.~\ref{boundaries}(b). In
Code K, the boundary is the smallest rectangle that encloses a
special curved contour \cite{Code K}. This curved contour is
produced by a 2D contour-plotting routine, and it is drawn not as
line segments around pixel edges but rather as a curve passing
through various pixels. Within a pixel, the pixel center is
assigned the value of the original pixel intensity, but other
points within a pixel are assumed to have other intensities, which
are calculated by 2D interpolation using four surrounding pixel
centers. Then, the contour-plotting routine draws a curve by
joining all points, with sub-pixel spacing, where the assumed
intensity is equal to the threshold, as shown in
Fig.~\ref{boundaries}(c) with a dash line. In both Codes A and K,
but not ImageJ, the boundary can enclose some pixels that are less
intense than the threshold.

In the second step, which is the same in all three codes we test,
the particle positions are calculated as the moment, i.e., as the
intensity-weighted position of pixels. The moment can be
calculated \cite{Samsonov:2000, Melandso:2000, Liu:2003} using
Eq.~(\ref{1stold}). However, we will find it better to use a
generalized form of the calculated particle position,
\begin{equation}\label{1stnew}
 {\bf X}_{calc}=\frac{\sum\limits_{k}{\bf X}_{k}(I_{k}-I_{base})}{\sum\limits_{k}(I_{k}-I_{base})},
\end{equation}
where the baseline value $I_{base}$ will be explained later. Note
that the calculated particle position depends on the selection of
pixels that are included in the summation in Eq.~(\ref{1stold}) or
Eq.~(\ref{1stnew}).

\section{METHOD}
\label{sec:method}

\subsection{Synthetic images}
\label{sec:syntheticimages}

To test methods of measuring particle positions, we calculate
position errors as compared to true positions in synthetic images.
For this purpose we cannot use actual experimental images because
the true position is generally not known. Synthetic images allow
us to vary the intensity and the size of a bright spot to find how
errors depend on these parameters.

Units used in this paper are pixel units for all distances
including ${\bf X}_{k}$, ${\bf X}_{calc}$, spot size and errors.
Intensities, including signal and noise, are specified in
intensity value units, i.e., a dimensionless integer ranging, for
example, from 0 to $2^{14}-1$ for a 14-bit camera.

We prepare synthetic images that resemble an experimental image
like Fig.~\ref{synthetic images}(a). The synthetic images have a
size of $64\times 64$ pixels, with one bright spot per image.
These images have three major attributes that we compute: the
spot's true position, the spatial profile of the signal, and the
noise.

First, the bright spot's true position is located near the image
center, but displaced in the $x$ and $y$ directions by a fraction
of a pixel. This is done using random numbers with a uniform
distribution (between 0 and 1) so that the true positions are
random and uniformly-distributed relative to pixel edges. Using
these random positions avoids any sampling bias.

Second, like other authors \cite{Crocker:1996, Gaussian profile},
we model the signal's spatial profile as a Gaussian
\begin{equation}\label{Gaussian}
 I_{sig}(x,y)=I_{peak}\ exp\left[-\frac{(x-x_{true})^2+(y-y_{true})^2}{r_{spot}^2}\right],
\end{equation}
characterized by the spot radius $r_{spot}$ and the peak intensity
$I_{peak}$. (This Gaussian is intended to approximate the actual
spatial profile, which depends on factors such as the particle
size, the camera's gamma, and lens defocusing.) To imitate the
collection of light onto a square pixel, we integrate this smooth
profile over each pixel's area. This yields the value $I_{sig\ k}$
of the signal in pixel $k$,
\begin{equation}\label{pixel intensity}
 I_{sig\ k}=\int\limits_{x=k_x-0.5}^{k_x+0.5}\int\limits_{y=k_y-0.5}^{k_y+0.5}I_{sig} (x,y) \mathrm{d}x\,\mathrm{d}y,\
\end{equation}
where $k_x$ and $k_y$ are the coordinates of pixel $k$.
Equation~(\ref{pixel intensity}) combined with
Eq.~(\ref{Gaussian}) can be evaluated efficiently using the error
function {\it erf}. (After this step, each bright spot has the
same total signal intensity $\sum I_{sig\ k}$, which was typically
37\,707 corresponding to the brightest spot in the experimental
image Fig.~\ref{experiment images}(a). In the experiment, not
every bright spot has the same total signal intensity because some
particles are levitated slightly above or below the brightest part
of the horizontal laser sheet.)

Third, we calculate a noise value $I_{noise\ k}$ which is
different for each pixel $k$. To simulate the experiment,
$I_{noise\ k}$ is chosen as a random intensity from the noise
distribution of our digital camera, Fig.~\ref{histogram}, which is
centered at an average intensity $I_{bg}$ = 384. Finally, the
intensity $I_{k}$ in each pixel is calculated as the sum of the
intensities of the signal and noise or a saturation value
$I_{sat}$, whichever is smaller,
\begin{equation}\label{signalnoise}
I_{k}={\rm Min}[(I_{sig\ k}+I_{noise\ k}), \ I_{sat}].
\end{equation}
We use $I_{sat}=2^{14}-1$ to simulate the saturation intensity of
a real camera with 14-bit resolution. Finally, we round $I_{k}$ to
an integer because cameras produce integer values for the
intensity of each pixel. The result of this calculation is a TIFF
image like Fig.~\ref{synthetic images}(b) or \ref{synthetic
images}(c).

Here we only consider bright spots that are circular, as in
Eq.~(\ref{Gaussian}). Although we do not simulate them here, we
note that non-circular bright spots can be analyzed using the
moment method, and they do occur in some experiments. Elliptical
particles arise when using analog video cameras with a limited
horizontal resolution, or when particles move rapidly during the
exposure time. The latter effect can be diminished by rastering a
laser beam rather than dispersing it into a constant sheet.
Defocusing a lens can result in non-circular spots, as in
Sec.~VII.

\subsection{Errors in calculated particle positions}
\label{sec:errorsincalculatedparticlepositions}

In this paper, we are mainly interested in errors in calculated
particle positions. In addition to errors in particle position,
the experimenter may also be concerned with errors in velocities
and other quantities computed from particle positions, as
discussed in the Appendix.

To characterize the error in calculated particle positions, we use
two diagnostics. First, we calculate sub-pixel maps, as described
in Sec.~II. Examining these sub-pixel maps qualitatively will
reveal pixel locking, which is one source of error. Second, we
characterize the total error, including both random errors and
pixel locking, as the root-mean-square (rms) difference of true
and calculated positions, i.e., the rms error
\begin{equation}\label{rms}
 \left [\frac{1}{N}\sum\limits_{m=1}^N(x_{m,calc}-x_{m,true})^2+(y_{m,calc}-y_{m,true})^2 \right ]^\frac{1}{2},
\end{equation}
where $m$ and $N$ are the index and total number, respectively, of
bright spots. While we can calculate the total error using
Eq.~(\ref{rms}), we cannot separately calculate the contributions
from random errors and pixel locking.

To achieve good statistics, we prepared over 370~\,000 synthetic
images, each with one bright spot. We used $N$ = 5000 when
calculating the rms error, and $N$~ = ~100~\,000 when calculating
sub-pixel maps. All of these images have different random true
positions for their bright spots, and the noise in each pixel is
different in all images.

\subsection{Parameters}
\label{sec:parameters}

To find a procedure for calculating position with minimal total
error, we will test three different codes, and we will vary
parameters corresponding to software and hardware adjustments that
an experimenter can make. We will now list these adjustments. The
experimenter can choose to focus the camera lens sharply, or
defocus it to make the bright spots in the image appear larger and
fill more pixels. As a second parameter, the experimenter can
adjust the image intensity by varying the camera aperture,
exposure time or illumination brightness. After recording images
with the camera, the experimenter will then use software. Here, we
test three moment method codes, as explained in Sec.~III. After
choosing a code, the experimenter can usually adjust two
parameters in that code: the threshold used in the first step, and
the baseline (if any) that is subtracted in the second step, as in
Eq.~(\ref{1stnew}).

Thus, we are motivated to analyze the impact of the following four
parameters that the experimenter must choose: focus, intensity,
threshold, and baseline. We do this by varying the values of
$r_{spot}$ (keeping the total signal intensity $\sum I_{sig\ k}$
as constant, as explained later), $I_{peak}$, $I_{th}$, and
$I_{base}$, respectively. We will vary each of these four
parameters in Sec.~V. We will also compare results from the three
different codes. The outcome of this analysis will be a practical
procedure, presented in Sec.~VI, that the experimenter can use to
minimize errors in calculated positions.

\section{RESULTS}
\label{sec:result}

\subsection{Threshold}
\label{sec:threshold}

The first parameter we vary is the threshold. The experimenter
will first choose a coarse range of threshold so that it is not so
low that noise is wrongly identified as particles and not so high
that fainter particles are overlooked. Then, within this coarse
range, a fine adjustment can be made to reduce error. Here, we
consider the fine adjustment.

Our results in Fig.~\ref{threshold} show that the total error
generally increases with threshold, and it also depends on the
choice of a code. We calculate the total error as the rms error,
using $N$ = 5000 images and Eq.~(\ref{rms}). Recall that the total
error includes both random and pixel-locking errors. The total
error generally increases with the threshold because raising the
threshold can eliminate pixels that have useful signal.

The total error exhibits not only a general increase with
threshold, but also an oscillation. This is seen in
Fig.~\ref{threshold}, where there are several oscillations
superimposed on the general trend. We cannot dismiss these
oscillations as mere statistical fluctuations because we achieved
good statistics by using 5000 particle positions. To identify the
cause of these oscillations, we tested how the boundaries that are
selected in the first step depend on the threshold. The result of
this test is shown in Fig.~\ref{boundary effect} as a table of the
boundaries selected by ImageJ. When the threshold is increased
slightly so that the boundary shrinks by one pixel, there is a
discrete jump in the calculated particle position. As the
threshold increases, there is a sequence of jumps, as the boundary
becomes smaller, one pixel at a time. These jumps, in aggregate
for many particles, lead to oscillations in the rms error as the
threshold is varied, which is the phenomenon we term the
``boundary effect''.

To identify the role of pixel locking in the total error, we
examine sub-pixel maps in Fig.~\ref{4sub-pixel maps}, which reveal
the importance of the threshold. For ImageJ, we provide sub-pixel
maps, Fig.~\ref{4sub-pixel maps}(a) and \ref{4sub-pixel maps}(b),
that correspond to the two thresholds that yielded the minimum and
maximum rms errors, respectively, in Fig.~\ref{threshold}. We note
that the signature of pixel locking is weaker, i.e., the sub-pixel
map is more uniform, for the case of the low threshold,
Fig.~\ref{4sub-pixel maps}(a), that yields the lowest total error.
Conversely, the signature of pixel locking is stronger, i.e., the
sub-pixel map has strongly non-uniform features, for the higher
threshold, Fig.~\ref{4sub-pixel maps}(b). In general, reducing the
threshold will reduce pixel locking. Other codes exhibit the same
trend, but with a different appearance for the sub-pixel maps, as
in Fig.~\ref{4sub-pixel maps}(c) and \ref{4sub-pixel maps}(d).

\subsection{Spot radius}
\label{sec:spotradius}

To simulate an experimenter's slight defocusing of a camera lens,
we varied the spot radius $r_{spot}$ in Fig.~\ref{radius}. We used
the Gaussian profile of Eq.~(\ref{Gaussian}), keeping the spot's
total signal intensity ($\sum I_{sig\ k}$ summed over all pixels)
constant. In this way we mimic an experiment where a particle
scatters the same finite number of photons into a camera lens
regardless of how the lens is focused. (We did not simulate the
ring-shaped bright spot that can occur for extreme defocusing.)
Defocusing can happen when an experimenter purposefully chooses to
defocus the lens for example to avoid saturating pixels; in other
cases, defocusing is not intentional but instead simply
unavoidable because particles are at different depths, as for
example in colloidal suspensions \cite{Kvarnstrom thesis} and 3D
dusty plasma suspensions \cite{Melzer:2006}.

Defocusing a lens during the experiment can actually be desirable.
By distributing the signal over a larger number of pixels, the
impact of a single pixel in the calculation of the particle's
position is less, so that pixel locking becomes weaker. On the
other hand, defocusing can reduce the signal in each pixel, so
that the signal-to-noise ratio (SNR) in each pixel becomes worse.
In other words, there can be a trade-off: defocusing can improve
pixel locking at the expense of making random errors worse. In our
results below we investigate this effect.

We should mention that when discussing defocusing, we always refer
to the experimenter's adjustment to the hardware when recording an
image. Unlike some other methods \cite{Crocker:1996}, here we do
not blur an image in software after it has been recorded by the
hardware.

The result in Fig.~\ref{radius} reveals three ranges of the spot
radius, where the second range is the most desirable. In the first
range, with small spot radii ($r_{spot} < 0.8$), the total error
diminishes with radius because the spot includes a saturated
pixel. Saturated pixels are undesirable because they introduce
wrong information for intensity into Eq.~(\ref{1stnew}). In the
second range, with slightly larger spot radii ($0.8 \le r_{spot}
\le 2.0$), the total error is smallest. In the third range, with
large spot radii ($r_{spot} > 2.0$), the total error generally
increases with $r_{spot}$ because the trade-off results in the
undesirable outcome of the worsened SNR in each pixel having a
stronger effect than the improved pixel locking due to defocusing.
The optimal spot radius is somewhere in the second range, which
for our parameters is approximately 0.8 - 2.0. We should
emphasize, however, that this range will vary depending on the
experiment due to different cameras (with different noise levels,
sensitivities and saturation levels), particle size, illumination,
and working distance between particles and lens. If the camera had
a higher noise level, the errors in this third range would be
larger and the experimenter would be unable to use much
defocusing. On the other hand, if the illumination were brighter,
then the entire curve in Fig.~\ref{radius} would shift toward
larger spot radii and the experimenter would be able to use more
defocusing.

In Fig.~\ref{radius} we also note an oscillation, superimposed on
the general trend, for $0.8 \le r_{spot} \le 2.0$. We attribute
this oscillation, which was observed previously in experiments by
K\"{a}ding and Melzer \cite{Melzer:2006}, to a boundary effect
similar to the one described above.

\subsection{Intensity}
\label{sec:intensity}

To simulate adjusting the illumination brightness, the exposure
time, or the camera aperture, we varied $I_{peak}$ in
Fig.~\ref{intensity}. As a result, the total signal intensity
$\sum I_{sig\ k}$ is varied, while $r_{spot}$ is kept constant. We
note that ImageJ yields the smallest total error.

The trend that would be expected for random errors only is a
downward slope as the intensity is increased, due to an improving
SNR in each pixel. This trend is indeed observed
Fig.~\ref{intensity}, but only for some of the data, as indicated
by solid curves. The opposite trend is also observed in
Fig.~\ref{intensity}, as indicated by dashed curves; since this
trend is opposite to what is expected for random errors only, we
attribute it to pixel locking. We term this particular effect of
pixel locking the ``pedestal effect.''

\subsection{Baseline}
\label{sec:baseline}

The pedestal effect is the result of a non-optimal choice of the
baseline. To illustrate this effect, in Fig.~\ref{pedestal} we
have sketched the cross section of a bright spot. The portion of
this cross section that lies within the boundary, defined by the
threshold, is shown shaded. This portion is divided in
Fig.~\ref{pedestal} into two parts, above and below the threshold.
We term the part below the threshold the ``pedestal,''
Fig.~\ref{pedestal}. The contribution of the pedestal to the
moment in Eq.~(\ref{1stnew}) can be large, or small, depending on
whether $I_{base}$ is small or large, respectively. In the extreme
case of a very large pedestal that dominates the calculation of
the particle position, the calculated particle position will often
fall near a pixel edge or midpoint, as it does in the case of a
centroid, thereby contributing to severe pixel locking. We term
this tendency toward severe pixel locking the ``pedestal effect.''
Below, we will determine the best choice of $I_{base}$ in order to
reduce the pedestal effect and the pixel-locking errors that it
introduces to the calculated particle positions.

To test the effect of the baseline that is chosen, in
Fig.~\ref{new intensity} we present the total error, calculated as
the rms error, for three different baseline values. From
Fig.~\ref{new intensity}, we see that the total error is reduced
by using a larger baseline value. The best choice is $I_{base} =
I_{th}$, because this results in the smallest total error. It also
minimizes pixel locking; the downward slope in Fig.~\ref{new
intensity} indicates that random errors dominate.

Thus, we conclude that in the second step, when using
Eq.~(\ref{1stnew}), the baseline should be chosen to be the same
as the threshold that was used in the first step. This can be done
most simply by subtracting the same threshold for every pixel in
the image. Alternatively, a different baseline level $I_{base\ k}$
for each pixel could be subtracted in Eq.~(\ref{1stnew}), to
account for a different background level for each pixel. The
latter method is useful because it allows the experimenter to
eliminate optical reflections due to room lights, for example. The
experimenter can calculate all the $I_{base\ k}$ baseline values
for the pixels as follows. First, the experimenter will use the
camera to record a ``dark-field'' image, with the illumination
turned off so that particles are not visible. To improve the
statistics, the experimenter can record a series of dark-field
images and average them, pixel-by-pixel, to reduce the effect of
random noise. This will yield an intensity $I_{dark\ k}$ for each
pixel.  Second, the baseline for each pixel will be calculated as
\begin{equation}\label{diffbase}
 I_{base\ k}=I_{dark\ k}+(I_{th}-I_{bg}).
\end{equation}
Here, $I_{bg}$ can be calculated as the average of  $I_{dark\ k}$
for pixels in the image.

With an optimal choice of both threshold and baseline, one can
achieve a sub-pixel map that shows no evidence of pixel locking,
as seen in Fig.~\ref{2sub-pixel maps}(a). This map was prepared
using ImageJ, with a baseline equal to the threshold. This choice
of a baseline minimizes the total error, as we learned above. The
reason that choosing $I_{base} = I_{th}$ minimizes the total error
is now clear: it greatly reduces pixel locking, so that mainly
errors from random noise remain. To further demonstrate the
usefulness of choosing a baseline equal to the threshold, compare
Fig.~\ref{4sub-pixel maps}(a) to Fig.~\ref{2sub-pixel maps}(a).
The former figure, which was prepared similarly except with no
baseline subtraction, reveals some pixel locking, while the latter
does not.

An experimenter, when attempting to choose optimal parameters,
will be unable to calculate the rms error, as we have done in
Fig.~\ref{new intensity}, for example. This is because the true
positions of particles are generally unknown. The experimenter
can, however, calculate sub-pixel maps, such as
Fig.~\ref{2sub-pixel maps}, because these require only calculated
positions. Comparing Fig.~\ref{2sub-pixel maps}(a) and
\ref{2sub-pixel maps}(b), which were both calculated with
$I_{base} = I_{th}$, but with a different $I_{th}$, we see that
the signature of pixel locking depends on the threshold.

We now find our best result by varying the threshold, in
Fig.~\ref{new threshold}, to minimize the rms error. The threshold
is the last parameter to choose, assuming that the experimenter
has already: (1) established the illumination level, (2) chosen a
camera with a given noise level, (3) defocused the camera lens to
avoid saturating pixels, and (4) planned to use a baseline
$I_{base} = I_{th}$. Noting that the rms error in Fig.~\ref{new
threshold} has several minima, we identify an optimal threshold by
choosing the lowest minimum. This yields our best result, an rms
error of 0.017. These same parameters also virtually eliminate the
signature of pixel locking in Fig.~\ref{2sub-pixel maps}(a). An
experimenter can identify an optimal $I_{th}$ similarly, but
without calculating the rms error, by examining sub-pixel maps for
various values of  $I_{th}$, and among the maps with weak
pixel-locking signatures, choosing the one with the lowest value
of $I_{th}$.

\section{ PRACTICAL PROCEDURE}
\label{sec: practicalprocedure}

We present here a practical procedure for using the moment method
that minimizes the total error, including both random errors and
pixel locking. This practical procedure includes first the use of
hardware to record images and then the use of software to analyze
them. Our software uses the moment method with baseline
subtraction as we tested above; there are also other well-tested
analysis methods that experimenters may wish to consider
\cite{Crocker:1996, Kvarnstrom thesis}.

For the hardware that produces the image, one will choose a camera
and make adjustments to the intensity and lens focusing. Choosing
a camera with low noise will not only reduce random errors; it
will also allow the use of a lower threshold which can improve
pixel locking. In using the camera, the optimal choices of
intensity and lens defocusing must be considered together. The
intensity can be varied, for example, by adjusting the camera
aperture, exposure time, or illumination level. To achieve a high
SNR in each pixel, we adjust the intensity upward as high as
possible without saturating pixels. Another way to improve SNR is
pixel binning, which also increase frame rate, but at the expense
of spatial resolution \cite{pixel binning}. If additional
intensity is available but pixels are saturated, the experimenter
can defocus the lens to avoid saturating the brightest pixels.
Defocusing the lens helps reduce pixel locking, but it can
increase random errors by reducing the SNR in each pixel;
therefore, defocusing beyond a certain point actually worsens the
total error. The optimal lens defocusing will depend on parameters
such as intensity, camera noise level, and number of camera bits,
which vary from one experiment to another. For the parameters we
simulated (see Fig.~\ref{radius}) we found that the optimal spot
radius was in the range 0.8 - 2.0, measured as the Gaussian
half-width. For other parameters, we can offer this general
guidance: the optimal lens defocusing will be determined by the
need to achieve an adequate SNR in each pixel. Noisier cameras or
weaker illumination will require less defocusing, while low-noise
cameras and brighter illumination will allow more defocusing. The
lens should generally be defocused at least enough to avoid
saturating pixels.

For the image analysis software, there are usually three important
choices. First, we prefer a code that has as its first step the
selection of a boundary that includes only contiguous pixels above
a threshold. The freely available ImageJ code selects such a
boundary. Second, if the boundary is selected as described above
in the first step, then in the second step, using
Eq.~(\ref{1stnew}), the baseline should be chosen equal to the
threshold, in order to reduce pixel locking. This can be done
either by subtracting the same baseline value from every pixel in
a single step, or by using Eq.~(\ref{diffbase}) with dark-field
images if the experimenter wishes to remove the effect of optical
reflections for example. Third, the threshold should be chosen in
a two-part process. To start, the experimenter should count the
number of particles that are identified, and then choose a coarse
range as explained in Sec.~V ~A. Next, within this coarse range,
sub-pixel maps should be calculated for various thresholds. In
order to reduce both random and pixel-locking errors, the user
should choose the lowest threshold that has a weak signature of
pixel locking.

The moment method can achieve very low errors in particle position
measurement when it is used optimally. For the case we simulated,
an rms error as small as 0.017 is achievable by making optimal
choices in the software. Even smaller errors could be attained if
the intensity were brighter or the camera had less noise.

Readers who wish to perform tests similar to ours may use our
codes and images \cite{our resource}.

\section{EXPERIMENTAL DEMONSTRATION}
\label{sec:experimentaldemonstration}

To demonstrate the practical procedure above, we used it in an
experiment. The results presented above, based on synthetic
images, indicate that both total errors and pixel locking will be
reduced if we follow the practical procedure. Using experimental
images, one can detect the signature of pixel locking using
sub-pixel maps. We describe next the hardware and software
components of our experimental test.

For the hardware, the experiment was similar to the one for
Fig.~\ref{experiment images}(a), including using the same 14-bit
camera, except that we improved the experimental method by
slightly defocusing the lens. A cropped portion of the $800\times
600$ pixels image Fig.~\ref{new exp images}(a) and a magnified
view Fig.~\ref{new exp images}(b) show that a bright spot fills
more pixels than in Fig.~\ref{experiment images}(b) where the lens
was sharply focused. Due to defocusing, the spots are slightly
noncircular. Additionally, we binned $2\times 2$ pixels. As a
result of these changes, the total intensity of a bright spot is
typically 39\,240, as compared to 21\,000 (with a maximum of
37\,707) for Fig.~\ref{experiment images}(a), and the noise peak
is shifted to a lower intensity. A further possible improvement in
the hardware is using a more powerful laser, and we plan to do
that in future experiments.

For the software, we used ImageJ to identify particles from 100
experimental images. We excluded any identified particles that
filled only one single pixel. First, we chose a coarse range for
the threshold by counting the number of identified particles as a
function of the threshold, Fig.~\ref{number vs threshold}. We
looked for a nearly flat portion, which is from 325 to 925 here,
and we chose that as the coarse range. Next, we calculated
particle positions using Eq.~(\ref{1stnew}), along with
Eq.~(\ref{diffbase}) to calculate $I_{base\ k}$ using an average
of 2000 dark-field images. We repeated these calculations of
particle positions for various thresholds, each time preparing a
sub-pixel map. Finally, we will examine these sub-pixel maps to
choose the lowest threshold that has a weak signature of pixel
locking.

In Fig.~\ref{8sub-pixel maps}, we present the sub-pixel map that
results from following our practical procedure in panel (a).
Examining this sub-pixel map, we see that it has no obvious
signature of pixel locking when viewed in its entirety. To search
for signatures, we zoom into the lower left corner,
Fig.~\ref{8sub-pixel maps}(b)-(h). There, we can identify an
artifact of pixel locking: a concentration of calculated positions
on pixel edges. Our practical procedure requires choosing the
lowest threshold with a weak signature of pixel locking. For our
results in Fig.~\ref{8sub-pixel maps}, thresholds in the range 325
- 425 have no identifiable signature, leading us to choose 325.

We conclude that the signature of pixel locking is vastly improved
by using our practical procedure. This conclusion is based on a
comparison of the sub-pixel maps in Fig.~\ref{8sub-pixel maps}(a)
and Fig.~\ref{sub-pixel map}(c). The latter was prepared for a
similar experiment but a different camera, illumination, and
analysis method.  The signature of pixel locking is profound in
Fig.~\ref{sub-pixel map}(c), but it is virtually undetectable in
Fig.~\ref{8sub-pixel maps}(a)-(c).

\section*{ACKNOWLEDGMENTS}

We thank O.~Arp and U.~Konopka for providing codes and helpful
discussions. We also thank R.~Mutel, V.~Nosenko, A.~Piel,
T.~Sheridan, and E.~Thomas~ Jr. for helpful discussions. This work
was supported by NASA and the U.S. Department of Energy.

\section*{APPENDIX: ERRORS IN OTHER QUANTITIES}

Errors in the calculated particle positions can introduce errors
in other quantities that are calculated from the positions. In
PTV, velocities are calculated as $v = (x_2-x_1)/ \Delta t$, as
discussed in Sec.~I. Pixel locking can affect the velocity
calculation greatly in experiments. For example, if pixel locking
is so severe that most calculated positions are located only at
pixel centers, then almost all particle velocities calculated in
PTV will be quantized as an integer number of pixel widths per
frame. These errors in calculating velocities can propagate to
other calculations. Velocity distribution functions $f(v)$ can be
badly affected, with noticeable peaks \cite{Nosenko:2006} that are
signatures of pixel locking. However, we have found that wave
spectra and velocity correlation functions are not affected so
badly.

While it is beyond the scope of this paper to completely
characterize the errors in $v$ or $f(v)$, we can discuss the
contributions to the total error in $v$. For PTV, the rms error,
${\delta v} = {(\overline{({\delta x_1}^2+{\delta x_2}^2-2 \delta
x_1 \delta x_2)}/ {\Delta t}^2)}^{ \frac{1}{2}}$, has two
contributions, $(\overline{{\delta x_1}^2+{\delta x_2}^2})/
{\Delta t}^2$ arising from the errors in position, and $(-2
\overline{ \delta x_1 \delta x_2})/ {\Delta t}^2$ arising from
correlations in the two errors. If the calculated position had
random errors only, the correlation $\overline{ \delta x_1 \delta
x_2}$ would be zero and the rms error in $v$ would be minimized
when the rms error in $x$ is minimized. However, pixel-locking
errors can have correlations, which will vary depending on the
velocities, and these will affect ${\delta v}$ in a way that is
difficult to predict.

Aside from these quantities, which are calculated from velocities,
experimenters often calculate other quantities from the position
itself. The mean-square displacement (MSD), which is used to
measure diffusion, is calculated from position. Particle position
errors can cause the MSD to be exaggerated significantly at small
times when the displacement is small, but not at large times when
the displacement is large \cite{Selle:2004}. Another use of
particle positions is the study of structure \cite{ Knapek:2007,
Quinn:1996}. While we have not analyzed the sensitivity of
structural analysis methods to particle position errors, we expect
that calculations that are sensitive to small changes in
interparticle distances, such as Voronoi maps for detecting
defects, will be more affected than correlation functions that use
data over a wide range of distances.

\begin{figure}[p]
\caption{\label{experiment images}Experimental bit-map images of a
monolayer suspension of microspheres in a dusty plasma. Each
bright spot corresponds to one particle. Here, (a) is 1/12 of the
original image from a digital camera and (b) is a magnified view,
showing that a bright spot fills several pixels, while in (c) from
an analog camera a bright spot fills about $5\times 5$ pixels.
Spot size depends on such factors as camera type and focusing. A
particle's position is calculated as the bright spot's center;
errors in this calculation are the topic of this paper. }
\end{figure}

\begin{figure}[p]
\caption{\label{histogram}Histogram of intensity values of pixels
in the original experimental image of Fig.~\ref{experiment
images}(a). The inset shows the same data with a logarithmic
scale. The prominent peak, centered at $I_{bg}$, is due to noise
in the camera. }
\end{figure}

\begin{figure}[p]
\caption{\label{sub-pixel map}Illustration of the method for
calculating a sub-pixel map. First, a $10\times 10$ pixel bit-map
image (not shown here) is analyzed to yield a map (a) of particle
positions. Second, the same positions are plotted relative to
pixel edges in (b); these values are the fraction parts of the
calculated positions. (c) An example sub-pixel map of $N$ = 617
particles, calculated from an experimental image (full view of
Fig.~\ref{experiment images}(c)), reveals pixel locking as a
tendency of calculated positions to be concentrated at favored
positions including the center and edges of pixels. }
\end{figure}

\begin{figure}[p]
\caption{\label{boundaries}Illustration of boundaries. In
algorithms for calculating particle positions from a bit-map
image, the first step is selecting the contiguous pixels to be
used, as defined by a boundary (solid white line) that encloses
them. The codes tested here differ only in the way they select
boundaries. (a) In ImageJ, only contiguous pixels above a
threshold are included in the boundary. Code A (b) and Code K (c)
use boundaries that are the smallest rectangles that enclose: all
the contiguous pixels above the threshold in Code A, or the dashed
contour produced by a 2D contour-plotting routine in Code K. }
\end{figure}

\begin{figure}[p]
\caption{\label{synthetic images}Magnified images of bright spots.
(a) Experimental image from a digital video camera. (b),(c)
Synthetic images, with a Gaussian profile centered on a known true
position, here with two different spot radii. In generating
synthetic images, we first choose the true position randomly, and
then calculate the intensity of each pixel using
Eq.~(\ref{signalnoise}) so that it includes both signal and noise.
}
\end{figure}

\begin{figure}[p]
\caption{\label{threshold}The rms error of calculated positions as
a function of the threshold $I_{th}$. In general, errors increase
with threshold, and superimposed on this increase is an
oscillation. The rms errors are always calculated as in
Eq.~(\ref{rms}) using $N = 5000$. (Here, $r_{spot} = 1.5$ pixel
units, $I_{peak} = 5334$ intensity value units, corresponding to a
total signal intensity $\sum I_{sig\ k} = 37\,707$. Also,
$I_{base} = 0$.) }
\end{figure}

\begin{figure}[p]
\caption{\label{boundary effect}Cause of oscillations. Boundaries,
selected in the first step of ImageJ, enclose fewer pixels as the
threshold is increased. Removing one pixel from the boundary
causes a discrete jump in the calculated particle position in
Eq.~(\ref{1stnew}). As the threshold increases, there is a
sequence of jumps, as the boundary becomes smaller, one pixel at a
time. These jumps, in aggregate for many particles, lead to
oscillations in the rms error as the threshold is varied, a
phenomenon we term the boundary effect. The three columns
correspond to three different true positions. }
\end{figure}

\begin{figure}[p]
\caption{\label{4sub-pixel maps}Sub-pixel maps for $N =
~100~\,000$ randomly distributed true positions. The signature of
pixel locking is generally more severe for higher thresholds.
(Here, $r_{spot} = 1.5$, $I_{peak} = 5334$, and $I_{base} = 0$.)}
\end{figure}

\begin{figure}[p]
\caption{\label{radius}Simulation of slight lens defocusing. The
optimal range of spot size lies between two other ranges: for very
small $r_{spot}$, errors worsen due to pixel saturation; for very
large $r_{spot}$, they worsen due to random errors. For our
parameters, these two ranges are for $r_{spot} < 0.8$ and
$r_{spot} > 2.0$, respectively. Oscillations in the optimal range
arise from a boundary effect. (Here, $I_{th} = 1000$, $I_{base} =
0$, and $\sum I_{sig\ k} = 37\,707$.) }
\end{figure}

\begin{figure}[p]
\caption{\label{intensity}The rms error as the intensity is
varied, to simulate adjusting the illumination brightness, the
exposure time or the camera aperture. The main trend is that the
error decreases with increasing intensity due to an improved
signal-to-noise ratio (SNR), as indicated by solid curves; the
opposite trend, indicated by dashed curves, is attributed to a
pixel-locking effect that we term the pedestal effect. (Here,
$r_{spot} = 1.5$, $I_{th} = 740$, and $I_{base} = 0$.) }
\end{figure}

\begin{figure}[p]
\caption{\label{pedestal}Cross section of a bright spot,
illustrating the ``pedestal.'' Pixels brighter than the threshold
identify the boundary for ImageJ in the first step. In the second
step, both shaded portions contribute to the calculated particle
position if $I_{base} = 0$, i.e., if no baseline is subtracted in
Eq.~(\ref{1stnew}). The lower shaded portion, marked ``pedestal,''
can heavily influence the calculated particle position. The
pedestal can be reduced by choosing $I_{base} = I_{bg}$, or
eliminated altogether by choosing $I_{base} = I_{th}$. }
\end{figure}

\begin{figure}[p]
\caption{\label{new intensity}Test of different baselines. The
best choice to minimize rms error is subtracting a baseline equal
to the threshold $I_{th}$ in Eq.~(\ref{1stnew}). (We used ImageJ,
and $r_{spot} = 1.5$, $I_{th} = 740$, and $I_{bg} = 384$.) }
\end{figure}

\begin{figure}[p]
\caption{\label{2sub-pixel maps}Sub-pixel maps, using a baseline
$I_{base} =I_{th}$ for two different thresholds (a) $I_{th} =
1150$ and (b) $I_{th} = 2950$. Comparing these panels shows that
the signature of pixel locking can be virtually eliminated, as in
(a), by making the best choice of threshold as well as choosing
$I_{base} = I_{th}$. (Here, we used the same 100~\,000 images as
in Fig.~\ref{4sub-pixel maps}.)}
\end{figure}

\begin{figure}[p]
\caption{\label{new threshold}Total error, using a baseline
$I_{base} = I_{th}$. Comparing to Fig. 6 where $I_{base} = 0$,
errors have been reduced. The lowest rms error that can be
achieved with these images is 0.017, using the same optimal choice
of parameters as in Fig.~\ref{2sub-pixel maps}(a). (We used the
same 5000 images as in Fig.~\ref{threshold}. Here and in
Fig.~\ref{2sub-pixel maps}, we used ImageJ.)}
\end{figure}

\begin{figure}[p]
\caption{\label{new exp images}Experimental bit-map images of a
monolayer suspension of microspheres in a dusty plasma. Here, (a)
is $1/12$ of the original image and (b) is a magnified view. A
bright spot fills about $5\times 5$ pixels. Compared to
Fig.~\ref{experiment images}(a), the hardware was improved by
slight lens defocusing.}
\end{figure}

\begin{figure}[p]
\caption{\label{number vs threshold}Choosing the coarse range of
threshold using experimental images. Counting the particles
identified in 100 images, we choose the nearly flat portion $325
\le I_{th} \le 925$ as the coarse range. Outside this coarse
range, many false particles appear at lower $I_{th}$ due to noise,
while many true particles are missed at higher $I_{th}$. Labels
a-h identify thresholds used in Fig.~\ref{8sub-pixel maps}.}
\end{figure}

\begin{figure}[p]
\caption{\label{8sub-pixel maps}Experimental sub-pixel maps for
different thresholds within the coarse range. Here, (a) is an
entire map, and (b)-(h) show the lower left corner. We choose the
lowest $I_{th}$ with a weak signature of pixel locking, 325. The
signature is stronger for $I_{th} \ge 525$, with a concentration
of calculated positions on pixel edges. Vastly better than
Fig.~\ref{sub-pixel map}(c), there is no obvious signature of
pixel locking for $I_{th} < 525$. (Here, we used ImageJ with
$I_{base\ k}$ calculated from Eq.~(\ref{diffbase}) and a
dark-field image.)}
\end{figure}


\begin{references}

\bibitem{Samsonov:2000}
D.~Samsonov, J.~Goree, H.~M.~Thomas, and G.~E.~Morfill, Phys. Rev.
E {\bf 61}, 5557 (2000).

\bibitem{Ivanov:2007}
Y.~Ivanov and A.~Melzer, Rev. Sci. Instrum. {\bf 78}, 033506
(2007).

\bibitem{Crocker:1996}
J.~C.~Crocker and D.~G.~Grier, J. Colloid Interf. Sci. {\bf 179},
298 (1996).

\bibitem{Kvarnstrom thesis}
M.~Kvarnstr\"{o}m and C.~A.~Glasbey, Biometrical J. {\bf 49}, 300
(2007).

\bibitem{Chang:1998}
K.~-A.~Chang and P.~L.~-F.~Liu, Phys. Fluids {\bf 10}, 327 (1998).

\bibitem{Selle:2004}
C.~Selle, F.~R\"{u}ckerl, D.~S.~Martin, M.~B.~Forstner, and
J.~A.~K\"{a}s, Phys. Chem. Chem. Phys.  {\bf 6}, 5535 (2004).

\bibitem{Shimizu:2005}
M.~Shimizu and M.~Okutomi, Int. J. Comput. Vision {\bf 63}, 207
(2005).

\bibitem{aperture book}
F.~A.~Jenkins and H.~E.~White, {\it Fundamentals of Optics}, 3rd
ed. (Mcgraw-Hill, New York, 1957), p. 302.

\bibitem{Melandso:2000}
F.~Melands\o{}, \AA.~Bjerkmo, G.~Morfill, H.~Thomas, and M.~Zuzic,
Phys. Plasmas {\bf 7}, 4368 (2000).

\bibitem{Liu:2003}
Bin~Liu, J.~Goree, V.~Nosenko, and L.~Boufendi, Phys. Plasmas {\bf
10}, 9 (2003).

\bibitem{ImageJ}
W.~S.~Rasband, computer code ImageJ version 1.34 (U.~S.~National
Institutes of Health, Bethesda, Maryland, 2006)
http://rsb.info.nih.gov/ij/. The ``center of mass'' is displaced
1/2 pixel in both directions, as compared to our definition.

\bibitem{Gai:2001}
M.~Gai, D.~Carollo, M.~Delb\`{o}, M.~G.~Lattanzi, G.~Massone,
F.~Bertinetto, G.~Mana, and S.~Cesare, Astron. Astrophys. {\bf
367}, 362 (2001).

\bibitem{Westerweel:1997}
J.~Westerweel, Meas. Sci. Technol. {\bf 8}, 1379 (1997).

\bibitem{Nosenko:2006}
V.~Nosenko, J.~Goree, and A.~Piel, Phys. Plasmas {\bf 13}, 032106
(2006).

\bibitem{Willert:1991}
C.~E.~Willert and M.~Gharib, Exp. Fluids {\bf 10}, 181 (1991).

\bibitem{Prasad:1992}
A.~K.~Prasad, R.~J.~Adrian, C.~C.~Landreth, and P.~W.~Offutt, Exp.
Fluids {\bf 13}, 105 (1992).

\bibitem{Gui:2002}
L. Gui and S. T. Wereley, Exp. Fluids {\bf 32}, 506 (2002).

\bibitem{Christensen:2004}
K.~T.~Christensen, Exp. Fluids {\bf 36}, 484 (2004).

\bibitem{Angele:2005}
K.~P.~Angele and B.~Muhammad-Klingmann, Exp. Fluids {\bf 38}, 341
(2005).

\bibitem{Nabach:2005}
H.~Nobach and M.~Honkanen, Exp. Fluids {\bf 38}, 511 (2005).

\bibitem{Melzer:2006}
S.~K\"{a}ding and A.~Melzer, Phys. Plasmas {\bf 13}, 090701
(2006).

\bibitem{Stein:2006}
A.~Stein, A.~Huertas, and L.~Matthies, ``Attenuating Stereo
Pixel-Locking via Affine Window Adaptation,'' presented at the
IEEE International Conference on Robotics and Automation, May,
2006. http://www.ri.cmu.edu/pubs/pub\_5378.html

\bibitem{Nehab:2006}
D.~Nehab, S.~Rusinkiewiez, and J.~Davis, ``Improved sub-pixel
stereo correspondences through symmetric refinement,'' presented
at the Tenth IEEE International Conference on Computer Vision,
2005.
http://ieeexplore.ieee.org/xpls/abs\_all.jsp?arnumber=1541303

\bibitem{Code A}
O. Arp, computer code DETECT\_PARTICLES\_2D.m (IEAP,
Christian-Albrechts-Universit\"{a}t, D-24098 Kiel, Germany, 2006).

\bibitem{Code K}
U. Konopka, computer code SPIT (Max-Planck-Institut f\"{u}r
extraterrestrische Physik, D-85741 Garching, Germany, 2005).

\bibitem{Gaussian profile}
H.~Huang, D.~Dabiri, and M.~Gharib, Meas. Sci. Technol. {\bf 8},
1427 (1997).

\bibitem{pixel binning}
Z.~M.~Zhou, B.~Pain, and E.~R.~Fossum, IEEE T. Electron Dev. {\bf
44}, 1764 (1997).

\bibitem{our resource}
See EPAPS Document No. for codes and sample images used in this
paper. This document can be reached through a direct link in the
online article's HTML reference section or via the EPAPS homepage
(http://www.aip.org/pubservs/epaps.html).

\bibitem{Knapek:2007}
C.~A.~Knapek, A.~V.~Ivlev, B.~A.~Klumov, G.~E.~Morfill, and
D.~Samsonov, Phys. Rev. Lett. {\bf 98}, 015001 (2007).

\bibitem{Quinn:1996}
R.~A.~Quinn, C.~Cui, J.~Goree, J.~B.~Pieper, H.~Thomas, and
G.~E.~Morfill, Phys. Rev. E {\bf 53}, R2049 (1996).
\end{references}
\end{document}